\documentclass[12pt]{article}
\usepackage{amssymb,amsmath,epsfig}

\begin{document}

\title{\bf Dynamics of Anisotropic Collapsing Spheres in Einstein Gauss-Bonnet Gravity}
\author{G.
Abbas$^1$\thanks{ghulamabbas@ciitsahiwal.edu.pk} and M. Zubair$^2$ \thanks{drmzubair@ciitlahore.edu.pk}\\\\
$^1$ Department of Mathematics, COMSATS\\
Institute of Information Technology, Sahiwal-57000, Pakistan.\\
$^2$ Department of Mathematics, COMSATS\\
Institute of Information Technology, Lahore, Pakistan.}
\date{}
\maketitle
\begin{abstract}
This paper is devoted to investigate the dynamics of the self
gravitating adiabatic and anisotropic source in $5D$ Einstein
Gauss-Bonnet gravity. To this end, the source has been taken as
Tolman-Bondi model which preserve inhomogeneity in nature. The field
equations, Misner-Sharp mass and dynamical equations have formulated
in Einstein Gauss-Bonnet gravity in $5D$. The junction conditions
have been explored between the anisotropic source and vacuum
solution in Gauss-Bonnet gravity in detail. The Misner and Sharp
approach has been applied to define the proper time and radial
derivatives. Further, these helps to formulate general dynamical
equations. The equations show that the mass of the collapsing system
increases with the same amount as the effective radial pressure
increases. The dynamical system preserve retardation which implies
that system under-consideration goes to gravitational collapse.
\end{abstract}
\

\section{Introduction}
The dynamics of self-gravitating objects is the topic of great
interest in general theory of relativity (GR), which is modern
theory of gravity. This problem becomes source of inspiration for
researchers when the stellar objects remain in stable for the most
of time against the perturbation caused by the self gravitational
force of the massive objects. This process provides the information
to study structure formation of the gravitationally collapsing
objects. In the relativistic gravitational physics the dynamics of
the stars studied by the Chandrasekhar \cite{1} first time in 1964,
since then there has been growing interest to study the dynamics of
stars in this research direction.  This work was extended by the
Herrera et al.\cite{2,3} explicitly for spherically symmetric heat
conducting, isotropic/anisotropic and viscous fluids in the
framework of GR. Recently, Herrera et al.\cite{5} have investigated
the dynamics of the expansion-free fluids using first order
perturbation of metric components as well as matter variables.

Several properties of the fluid play a dominant role in dynamical
process of the gravitationally collapsing objects. Herrera et
al.\cite{7} have studied the expansion free condition for the
collapsing sphere. Herrera and his collaborators \cite{7f,7a}
discussed the dynamical process of gravitational collapse using
Misner and Sharp's formulation. They considered the matter
distribution with shear-free spherically symmetric. The realistic
model of heat conducting star which shows dissipation in the form of
heat flux in radial direction and shear viscosity was studied by
Chan \cite{7b}. Herrera et al. \cite{7c} also formulated the
dynamical equations of the fluids which contains heat flux,
radiation and bulk viscosity and then coupled these equations with
causal transport equations. The inertia of heat flux and its
significance in the dynamics of dissipative collapse was studied by
Herrera \cite{7d}. The present paper is the particular case
(adiabatic case) of this work in $5D$ Einstein Gauss-Bonnet gravity.
Sharif and Azam \cite{11a}-\cite{13} have studied the effects of
electromagnetic field on the dynamical stability of the collapsing
dissipative and non dissipative fluids in spherical and cylindrical
geometries. This work has been further extended by Sharif and his
collaborators \cite{15}-\cite{22} in modified theories of gravity,
like $f(R)$ and $f(T)$ and $f(R,T)$. Recently \cite{23}, Abbas and
Sarwar have studied the dynamical stability of collapsing star in
Gauss-Bonnet gravity.

The dynamical system dealing with the dimensions greater or equal to
5 are usually discussed in the Gauss-Bonnet gravity theory. The
natural appearance of this theory occurs in the low energy effective
action of the modern string theory. Boulware and Deser \cite{44}
investigated the black hole (BH) solutions in $N$ dimensional string
theory with four dimensional Gauss-Bonnet invariant. This work is
the extension of $N$ dimensional solutions formulated by Tangherili
\cite{45}, Merys and Perry \cite{46}. Wheeller \cite{47} discussed
the spherically symmetric BH solutions with their physical
properties in detail. The topological structure of nontrivial BHs
has been explored by Cai \cite{48}. Kobayashi \cite{49} and Maeda
\cite{50} have formulated the structure of Vaidya BH in Gauss-Bonnet
gravity. All investigations show that the presence of the
Gauss-Bonnet term in the field equations would effect the final
state of the gravitational collapse. Recently \cite{51}, Jhinag and
Ghosh have consider the $5D$ action with the Gauss-Bonnet terms in
Tolman-Bondi model and give an exact model of the gravitational
collapse of a inhomogeneous dust. Motivated by these studies, we
have studies, we have explored the dynamics of the gravitationally
collapsing spheres in Einstein Gauss-Bonnet gravity. This paper is
extension of Herrera work\cite{7d} in Einstein Gauss-Bonnet gravity.
We would like to mention that the objective of this paper is to
study the effects of Gauss-Bonnet term on dynamics of collapsing
system. As heat flux is absent in the source equation, so transport
equations and their coupling with the dynamical equations is not the
objective of this paper. Hopefully it will be discussed explicitly
elsewhere.

This paper is organized as follow: In section \textbf{{2}} the
Einstein Gauss-Bonnet field equations and matching conditions have
been discussed. The dynamical equations have been formulated in
section \textbf{3}. We summaries the results of the paper in the
last section.

\section{Einstein Gauss-Bonnet Field Equations }

Consider the following action in $5D$
\begin{equation}\label{1}
S=\int d^{5}x\sqrt{-g}\left[ \frac{1}{2k_{5}^{2}}\left( R+\alpha
L_{GB}\right) \right] +S_{matter}
\end{equation}%
where $R$ ia the Ricci scalar in $5D$ and $k^2_{5}={8\pi G_{5}}$ is
coupling constant in $5D$. Also, the Gauss-Bonnet Lagrangian has the
form
\begin{equation}\label{2}
L_{GB}=R^{2}-4R_{\alpha\beta}R^{\alpha\beta}+R_{\alpha\beta
cd}R^{\alpha\beta cd}
\end{equation}
where coefficient $\alpha$ is coupling constant in Einstein
Gauss-Bonnet gravity. Such action is derivable in the low-energy
limiting case of super-string theory. Here, $\alpha $ is treated as
the inverse of string tension which is positive definite and $\alpha
\geq 0$ in this paper. For the $4D$ manifold, Gauss-Bonnet terms do
not contribute to field equations. The variation of action (\ref{1})
with respect to $5D$ metric tensor yields the following set of field
equations
\begin{equation}\label{3}
{G}_{\alpha\beta}=G_{\alpha\beta}+\alpha
H_{\alpha\beta}=T_{\alpha\beta},
\end{equation}
where
\begin{equation}\label{4}
G_{\alpha\beta}=R_{\alpha\beta}-\frac{1}{2}g_{\alpha\beta}R
\end{equation}
is the Einstein tensor and
\begin{equation}\label{5}
H_{\alpha\beta}=2\left[ RR_{\alpha\beta}-2R_{a\alpha }R_{b}^{a
}-2R^{a b
}R_{a\alpha b\beta }+R_{a}^{a b \gamma }R_{b\alpha \beta \gamma }%
\right] -\frac{1}{2}g_{\alpha\beta}L_{GB},
\end{equation}
is the Lanczos tensor.

A spacelike $4D$ hypersurface $\Sigma^{(e)}$ is taken such that it
divides a $5D$ spacetime into two 5D manifolds, $M^-$ and $M^+$,
respectively. The $5D$ TB spacetime is taken as an interior manifold
$M^-$  which is inner region of a collapsing inhomogeneous and
anisotropic star is given by \cite{51}
\begin{equation}\label{6}
ds_{-}^2=-dt^2+B^2dr^2+C^2(d\theta^2+\sin^2{\theta}d\phi^2
+\sin^2{\theta}\sin^2{\phi}d\psi^2),
\end{equation}
where $B$ and $C$ are functions of $t$ and $r$. The energy-momentum
tensor $T_{\alpha \beta }^{-}$ for anisotropic fluid has the form
\begin{equation}\label{7}
T_{\alpha \beta }^{-}=(\mu +P_{\perp })V_{\alpha }V_{\beta
}+P_{\perp }g_{\alpha \beta }+(P_{r}-P_{\perp })\chi _{\alpha }\chi
_{\beta },
\end{equation}
where $\mu $ is the energy density, $P_{r}$ the radial pressure,
$P_{\perp }$the tangential pressure, $V^{\alpha }$ the four velocity
of the fluid and $ \chi _{\alpha }$ a unit four vector along the
radial direction. These
quantities satisfy,%
\begin{equation}
V^{\alpha }V_{\alpha }=-1\ \ ,\ \ \ \ \ \chi ^{\alpha }\chi _{\alpha
}=1\ \ ,\ \ \ \ \ \chi ^{\alpha }V_{\alpha }=0  \label{N8}
\end{equation}%
The expansion scalar $\Theta $ for the fluid is given by
\begin{equation}\label{8}
\Theta =V_{\ ;\ \alpha ,}^{\alpha }.
\end{equation}%
Since we assumed the metric (6) comoving, then
\begin{equation}\label{9}
V^{\alpha }=\delta _{0}^{\alpha }\ ,\ \ \ \ \ \chi ^{\alpha
}=B^{-1}\delta _{1}^{\alpha }\
\end{equation}%
and for the expansion scalar, we get
\begin{equation}\label{10}
\Theta =\frac{\dot{B}}{B}+\frac{3\dot{C}}{C}.
\end{equation}
Hence, Einstein Gauss-Bonnet field equations take the form%
\begin{eqnarray}\nonumber
{\kappa}^2_{5}\mu &&=\frac{12\left( C^{\prime 2}-B^{2}\left(
1+\dot{C}^{2}\right) \right) }{C^{3}B^{5}}\left[ C^{\prime
}B^{\prime }+B^{2}\dot{C}\dot{B}-BC^{\prime \prime }\right] \alpha
\\\label{11} &&\ \ \ \ \ -\frac{3}{B^{3}C^{2}}\left[ B^{3}\left(
1+\dot{C}^{2}\right) +B^{2}C\dot{C}\dot{B}+CC^{\prime }B^{\prime
}-B(CC^{\prime \prime }+C^{\prime 2})\right]\\\label{12a}
{\kappa}^2_{5}p_{r} &&=-12\alpha \left( \frac{1}{C^{3}}-\frac{C^{^{\prime }2}}{B^{2}C^{3}}%
+\frac{\dot{C}^{2}}{C^{3}}\right) \ddot{C}+3\frac{C^{^{\prime }2}}{B^{2}C^{2}%
}  -3\Big(\frac{1+\dot{C}^{2}+C\ddot{C}}{C^{2}}\Big)\\\nonumber
{\kappa}^2_{5}p_{\perp } &&=\frac{4\alpha }{B^{4}C^{2}}\Big[
-2B\left( B^{^{\prime
}}C^{^{\prime }}+B^{2}\dot{B}\dot{C}-BC^{^{\prime \prime }}\right) \ddot{C}%
 +B\left( C^{^{\prime
}2}-B^{2}\left( 1+\dot{C}^{2}\right) \right)
\ddot{B}\\\nonumber&&+2\Big( \dot{B}C^{^{\prime
}}-B\dot{C}^{^{\prime }}\Big] -\frac{1}{B^{3}C^{2}}\Big[ B^{3}\Big(
1+\dot{C}^{2}+2C\ddot{C}\Big) +B^{2}C\left(
2\dot{C}\dot{B}+C\ddot{B}\right)\\&&+2CC^{^{\prime }}B^{^{\prime
}}-2B\left( CC^{^{\prime \prime }}+C^{^{\prime }2}\right)\Big]
\label{13}\\
 &&\frac{12\alpha }{B^{5}C^{3}}\left( \dot{B}C^{^{\prime
}}-B\dot{C}^{^{\prime
}}\right) \left( B^{2}\left( 1+\dot{C}^{2}\right) -C^{^{\prime }2}\right) -3%
\frac{B\dot{C}^{^{\prime }}-\dot{B}C^{^{\prime }}}{B^{3}C}=0.
\label{14}
\end{eqnarray}
The mass function $m(t,r)$ analogous to Misner-Sharp mass in $n$
manifold without ${\Lambda}$ is given by \cite{51a}
\begin{equation}\label{15}
m(t,r)=\frac{(n-2)}{2k_{n}^{2}}{V^k}_{n-2}\left[ R^{n-3}\left(
k-g^{ab}R,_{a}R,_{b}\right) +(n-3)(n-4)\alpha \left(
k-g^{ab}R,_{a}R,_{b}\right) ^{2} \right],
\end{equation}
where a comma denotes partial differentiation and ${V^k}_{n-2}$ is
the surface of $(n-2)$ dimensional unit space. For $k=1$,
${V^1}_{n-2}=\frac{2{\pi}^{(n-1)/2}}{\Gamma((n-1)/2)}$, using this
relation with $n=5$ and Eq.(\ref{6}), the mass function (\ref{15})
reduces to
\begin{equation}\label{16}
m(r,t)=\frac{3}{2}\left[ C^{2}\left( 1-\frac{C^{^{\prime }2}}{B^{2}%
}+\dot{C}^{2}\right) +2\alpha \left( 1-\frac{C^{^{\prime }2}}{B^{2}}+\dot{C}%
^{2}\right) ^{2}\right]
\end{equation}

In the exterior region to $\Sigma^{(e)}$, we consider Einstein
Gauss-Bonnet Schwarzschild solution which is given by

\begin{equation}\label{c1}
ds_{+}^2=-F(R)d{\nu}^2-2d\nu dR+R^2(d\theta^2+\sin^2{\theta}d\phi^2
+\sin^2{\theta}\sin^2{\phi}d\psi^2),
\end{equation}
where
$F(R)=1+\frac{{R}^2}{4\alpha}-\frac{{R}^2}{4\alpha}\sqrt{1+\frac{16\alpha
M}{\pi {R}^4}}$.

The smooth matching of the $5D$ anisotropic fluid sphere (\ref{6})
to GB Schwarzschild BH solution (\ref{c1}), across the interface at
$r = {r_{\Sigma}}^{(e)}$ = constant, demands the continuity of the
line elements and extrinsic curvature components (i.e., Darmois
matching conditions \cite{53}), implying
\begin{eqnarray}\label{c2}
dt \overset{\Sigma^{(e)}}{=}\sqrt{F(R)}d\nu,\\
R \overset{\Sigma^{(e)}}{=}R, \\\label{cm}
m(r,t)\overset{\Sigma^{(e)}}{=}M,
\end{eqnarray}
\begin{eqnarray}\nonumber
 &&-12\alpha \left( \frac{1}{C^{3}}-\frac{C^{^{\prime }2}}{B^{2}C^{3}}%
+\frac{\dot{C}^{2}}{C^{3}}\right) \ddot{C}+3\frac{C^{^{\prime }2}}{B^{2}C^{2}%
}  -3\Big(\frac{1+\dot{C}^{2}+C\ddot{C}}{C^{2}}\Big)\\
 &&\overset{\Sigma^{(e)}}{=}\frac{12\alpha }{B^{5}C^{3}}\left( \dot{B}C^{^{\prime
}}-B\dot{C}^{^{\prime
}}\right) \left( B^{2}\left( 1+\dot{C}^{2}\right) -C^{^{\prime }2}\right) -3%
\frac{B\dot{C}^{^{\prime }}-\dot{B}C^{^{\prime }}}{B^{3}C}.
\label{c3}
\end{eqnarray}
Comparing Eq.(\ref{c3}) with (\ref{12a}) and (\ref{14}) (for detail
see \cite{12}), we get
\begin{equation}\label{c4}
p_r\overset{\Sigma^{(e)}}{=}0.
\end{equation}
Hence, the matching of the interior inhomogeneous anisotropic fluid
sphere (\ref{6}) with the exterior vacuum Einstein Gauss-Bonnet
spactime (\ref{c1}) produces Eqs.(\ref{6}) and (\ref{cm}).

\section{Dynamical Equations}
In this section, we formulate the equations that deal with the
dynamics of collapsing process in Einstein Gauss-Bonnet gravity.
Following Misner and Sharp formalism \cite{52}, we discuss the
dynamics of the collapsing system. We introduce proper time
derivative as well as the proper radial as follows:
\begin{equation}\label{D1}
D_{T}=\frac{\partial}{\partial{t}},\quad
D_{R}=\frac{1}{R'}\frac{\partial}{\partial{r}},\quad R=C.
\end{equation}
The velocity of the collapsing fluid is the proper time derivative
of $R$ defined as
\begin{equation}\label{D2}
U=D_{T}(R)\equiv \dot{C}.
\end{equation}
Using above result in the mass function given by Eq.(\ref{16})
\begin{equation}\label{D3}
m(r,t)=\frac{3}{2}\left[ C^{2}\left( 1-\frac{C^{^{\prime }2}}{B^{2}%
}+U^{2}\right) +2\alpha \left( 1-\frac{C^{^{\prime }2}}{B^{2}}+U%
^{2}\right) ^{2}\right].
\end{equation}
Solving above equation for $\frac{C^{^{\prime }}}{B}$, we get the
positive and negative roots, the positive roots are given by
\begin{equation}\label{D3}
E=\frac{C^{^{\prime
}}}{B}=\sqrt{1+U^2+\frac{R^2}{4\alpha}\pm\frac{\sqrt{3R^2+16m\alpha}}{4\sqrt{3}\alpha}}.
\end{equation}
The rate of change of mass (Eq.(\ref{16})) with respect proper time
is given by
\begin{equation}\label{D3}
D_{{T}}m(t,r)=-{\kappa_5}^2P_{r}U{R}^{3},
\end{equation}
where we have used Einstein Gauss-Bonnet field equations
Eqs.(\ref{12a}) and (\ref{14}). The right hand side of this equation
has a single term . The this term is due to effective pressure
(means the pressure is effected by the Gauss-Bonnet term) in
$r$-direction. This term is positive in case of collapse ($U<0$).
This implies that as effective pressure in $r$-direction increases,
mass (energy) also increases with the same amount. Similarly, we can
calculate
\begin{equation}\label{D4}
D_{R}m(t,r)=\frac{2}{3}k^2_{5}\mu R^{3},
\end{equation}
where we have used Einstein Gauss-Bonnet field equations
Eqs.(\ref{11}) and (\ref{14}). This equation explain how effective
energy density affects the mass between neighboring hypersurfaces in
the interior fluid distribution. Integration of Eq.(\ref{D4}) yields
\begin{equation}\label{D5}
m(t,r)=\frac{2}{3}k^2_{5}\int^{{R}}_{0}({R}^{3}\mu)d{R}.
\end{equation}
The dynamical equations can be obtained from the contracted Bianchi
identities ${T^{ab}}_{;b}=0$. Consider the following two equations
\begin{eqnarray}
{T^{\alpha\beta}}_{;\beta}V_{\alpha}&&=\left[ \dot{\mu}+\left( \mu
+P_{r}\right) \frac{\dot{B}}{B}+3\left( \mu +P_{\perp }\right)
\frac{\dot{C}}{C}\right] =0 , \label{D6}\\ T_{;\beta }^{\ \alpha
\beta }\chi _{\alpha }&&=\frac{1}{B}\left[
P_{r}^{^{\prime }}+3\left( P_{r}-P_{\perp }\right) \frac{C^{^{\prime }}}{C}%
\right] =0. \label{D7}
\end{eqnarray}
The acceleration of the collapsing fluid is defined as
\begin{equation}\label{D8}
D_{{T}}U=\ddot{C}.
\end{equation}
Using Eqs.(\ref{D7}), (\ref{D8}) and (\ref{12a}), we get
\begin{eqnarray}
&&\Big[12\alpha
C\Big(\frac{9(p_r-p_{\bot})^2}{C^3}(1+U^2)+(\frac{{p_r}'}{
B^2}+1)\Big)\Big]D_T
U\\\nonumber&&=-9(p_r-p_{\bot})^2\Big[\kappa^2_5p_rC+3(\frac{1}{C}+\frac{U^2}{C})\Big].\label{D9}
\end{eqnarray}
This equation yields the effect of different forces on the
collapsing process. It can be interpreted in the form of Newton's
second law of motion i.e., Force = mass density $\times$
acceleration. The term within square bracket on left side of above
equation represent the inertial or passive gravitational mass. All
the quantities on right side in square bracket are positive, hence
this side is consequently negative and implies the retardation of
dynamical system giving rise to the collapse of the system.

\section{Outlook}

This paper investigates the effects of the Gauss-Bonnet term on the
dynamics of anisotropic fluid collapse in the $5D$ Einstein
Gauss-Bonnet gravity. We have extended the work of Herrera \cite{7d}
to $5D$ Einstein Gauss-Bonnet gravity. To this end, the
non-conducting anisotropic fluid with $5D$ spherical symmetry has
been taken as the source of gravitation in Einstein Gauss-Bonnet
gravity. The Misner-Sharp mass has been calculated in the present
scenario. The smooth matching of the interior source has been
carried out with $5D$ Schwarzschild BH solution in Einstein
Gauss-Bonnet gravity by using the Darmois \cite{53} junction
conditions. The matching of the two regions implies the vanishing of
radial pressure over the boundary of the star and continuity of the
gravitational masses in the interior and exterior regions. By using
the Misner and Sharp approach for the proper time and radial
derivatives, we have formulated the velocity as well as acceleration
of the system. These definitions have been also applied to formulate
the general dynamical equations in Gauss-Bonnet gravity.

The analysis of the dynamical equations predicts the following
consequences:
\begin{itemize}
\item Mass of the collapsing spheres increases with the passage of
time.
\item Effective energy density of the system would effects the mass of the system during the different stages of the
collapse
\item The system under consideration goes to retardation implying the gravitational
collapse.
\end{itemize}

We would like to mention that transport equations and their coupling
with dynamical equations are not the objective of this paper as heat
flux is absent in anisotropic fluid. This will be done in an other
investigation with the \textbf{inclusion of charge term in the
interior source }in future.
 
\end{document}